# A hybrid optoelectronic Mott insulator

H. Navarro[1*], J. del Valle[1,2], Y. Kalcheim[1], N. M. Vargas[1], C. Adda[1], M-H. Lee[1], P. Lapa[1], A. Rivera-Calzada[3], I. Zaluzhnyy[1], E. Qiu[1], O. Shpyrko[1], M. Rozenberg[4], A. Frano[1] and Ivan K. Schuller[1]

[1] Department of Physics, Center for Advanced Nanoscience, University of California, San Diego, 92093, USA.

[2] Department of Quantum Matter Physics, University of Geneva, 24 Quai Ernest-Ansermet, Geneva, Switzerland.

[3] Departamento de Física de Materiales, Universidad Complutense de Madrid, 28040 Madrid, Spain.

[4] Laboratoire de Physique des Solides, CNRS, Université Paris Saclay, 91405 Orsay Cedex, France.

**Corresponding Author**
*E-mail: hnavarro@physics.ucsd.edu

**The coupling of electronic degrees of freedom in materials to create 'hybridized functionalities' is a holy grail of modern condensed matter physics that may produce novel mechanisms of control. Correlated electron systems often exhibit coupled degrees of freedom with a high degree of tunability which sometimes lead to hybridized functionalities based on external stimuli[1-3]. However, the mechanisms of tunability and the sensitivity to external stimuli are determined by intrinsic material properties which are not always controllable. A Mott metal-insulator transition[4], which is technologically attractive due to the large changes in resistance, can be tuned by doping[5], strain[6,7], electric fields[8,9], and orbital occupancy[10] but cannot be, in and of itself, controlled externally with light. Here we present a new approach to produce hybridized functionalities using a properly engineered photoconductor/strongly-correlated hybrid heterostructure, showing control of the Metal-to-Insulator transition (MIT) using optical means. This approach combines a photoconductor, which does not exhibit an MIT, with a strongly correlated oxide, which is not photoconducting. Due to the close proximity between the two materials, the heterostructure exhibits large volatile and nonvolatile, photoinduced resistivity changes and substantial photoinduced shifts in the MIT transition temperatures. This approach can potentially be extended to other judiciously chosen combinations of strongly correlated materials with systems which exhibit optically, electrically or magnetically controllable behavior.**

1. **Introduction**

The physical properties of strongly correlated materials appeal to a broad scientific community because of the versatility and tunability of their electronic responses via internal and/or external perturbations[1]. However, the number of ways of controlling a single correlated material is limited by the available internal degrees of freedom. This presents a difficult challenge when a device concept requires a specific mechanism of control that is not accessible within a material. Here we present a simple, general solution to this limitation by judiciously designing a heterostructure that hybridizes the functionalities of two seemingly unrelated materials: a strongly correlated Mott insulator with a photoconducting semiconductor. This gives us exquisite control of a metal-to-insulator transition (MIT) via optical means. With its striking simplicity, our methodology can be expanded to create hybridized functionalities across e.g. semiconductors, strongly correlated, magnetic, and topological materials.

The Mott MIT in vanadium oxides—a hallmark of strong correlations—exemplifies behavior that is attractive for its fundamental interest and potential applications in new technologies[2,11-17]. This transition is characterized by an abrupt change in the electrical resistivity by several orders of magnitude, and can be controlled by modulating the temperature, doping, pressure and electric field[4-6,8,10,18-20]. Some materials, such as $VO_2$, also feature large changes in their optical properties across the MIT, making them attractive for optoelectronic applications[21-23]. Earlier studies[7,24] used $VO_2$ to modify the boundary conditions in heterostructures and this way to implement electrically controlled optical devices. However, the reverse effect, i.e. controlling the electrical properties by applying light, remains elusive. Some studies have shown that visible and UV light can be used to tune the transition in $VO_2$[25], but the reported modulations are rather small. To increase the $VO_X$ sensitivity to light, we interface the vanadium oxide layer with a material with high light sensitivity, a semiconducting photoconductor. In this geometry (Figure 1), light can strongly affect the MIT through proximity effects, such as carrier doping and/or light-induced interfacial catalytic reactions. We used the archetypical photoconductor cadmium sulfide (CdS), one of the most studied chalcogenide semiconductor materials[26] with a high concentration of donor carriers[27]. When exposed to visible light, CdS changes its conductivity substantially by generating electron and hole carriers[28] with electrons as the main carriers in the photoexcited state. Moreover, when the CdS interacts with oxygen and light, this material can behave as a photocatalyst[29,30]. In this work, we have grown photoconducting CdS thin films on $VO_X$ layers (see Figure 1) in order to produce a photosensitive Mott insulator. The thickness of the VO$_X$ was reduced as much as possible to maximize the effect of the interface on the transport properties. Our results show that such hybrid structures exhibit very large, volatile and nonvolatile photoinduced modulations of both the amplitude and the critical temperature of the MIT.

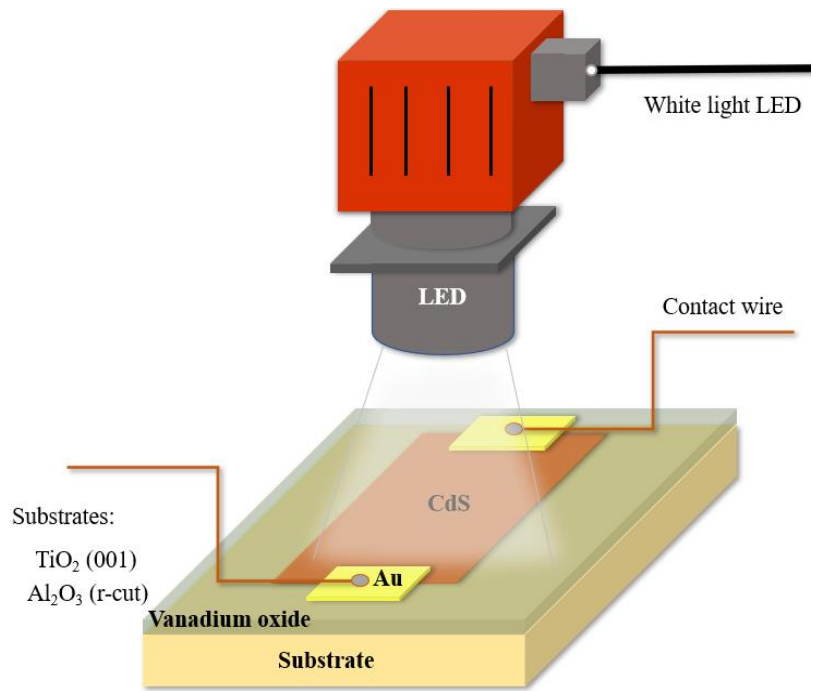

**Figure 1**: **Schematic device and experimental setup.** A CdS (80 ± 1 nm) film was deposited over vanadium oxide thin films (10 ± 1 nm both). The CdS/V$_2$O$_3$ and CdS/VO$_2$ heterostructures were illuminated with a white light LED. CdS/V$_2$O$_3$ heterostructures were directly grown on Al$_2$O$_3$ (R-cut) substrates, and CdS/VO$_2$ were grown on a TiO$_2$ (001) substrate for optimal growth. The bilayer conductivity was measured with two probes contacted by two Au pads.

2. **Results and discussion**

Figure 2 shows the hysteretic resistance of the samples as a function of temperature in dark, due to the first order MIT of over five orders of magnitude in CdS/V$_2$O$_3$ and three orders of magnitude in CdS/VO$_2$. This temperature-dependent behavior is identical to that of bare VO$_2$ and V$_2$O$_3$, shown in Figure 3, confirming that growing CdS over the vanadates does not modify their properties. Figure 2 shows the effect of illumination of these samples with varying light power densities. As the light power density is increased, the MIT temperature is reduced and the resistance of the insulating ground state decreases. The effects are most notable in the CdS/V$_2$O$_3$ bilayer, where the MIT is almost completely suppressed for a light power density of 731 mW/cm$^2$. This results in a resistance drop of more than 6 orders of magnitude compared to the case with no light. The light effects are smaller, though still significant, for VO$_2$ with T$_{MIT}$ decreasing by ~15 K when illuminated with a power of 731 mW/cm$^2$, meaning a drop of about half an order of magnitude in the resistance below the transition.

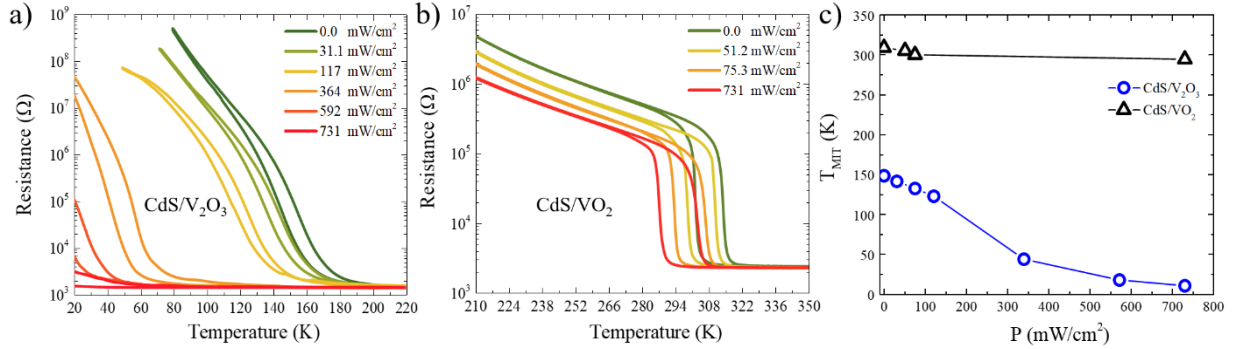

**Figure 2**: **Light-induced modification of the metal-insulator transition in CdS/Vanadium oxide heterostructures.** Electrical transport measurements of the resistance as a function of temperature from the bilayers (a) CdS/V$_2$O$_3$, (b) CdS/VO$_2$. (c) Values of metal-insulating transition temperatures (T$_{MIT}$) plotted as a function of the power density of the light source.

In the absence of CdS, the effect of light directly on the vanadates is much smaller, as shown in Figure 3. This implies that the CdS plays a crucial role in our observations and that sample heating do not play a role. To investigate further the origin of this effect, we repeated the photodoping experiments with thicker VO$_X$ films (100 ± 2 nm), which exhibit no changes in the T$_{MIT}$ (see Figure S.4). This points to a proximity effect localized at the CdS/VO$_X$ interface. To further corroborate this, we fabricated samples in which a 10-30 nm thick, insulating Al$_2$O$_3$ layer was sandwiched between the vanadium oxide and the CdS (see Figure S.5), so that any proximity effect would be suppressed. No MIT modification was observed in this case.

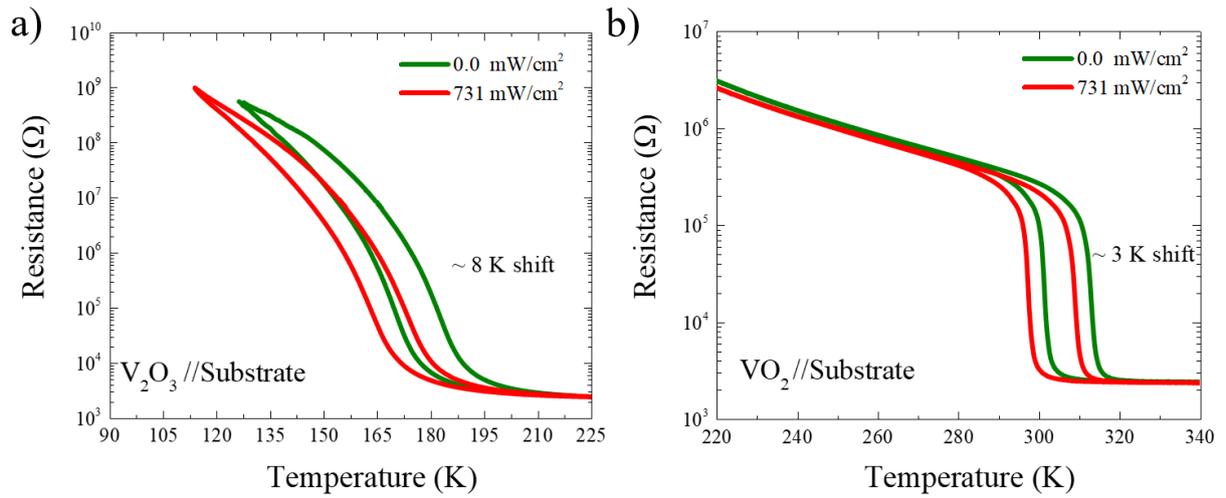

**Figure 3**: **Light-induced modulation of the metal-insulator transition in bare vanadium oxide films.** Electrical transport measurements of the resistance as a function of the temperature in bare vanadium oxide thin film without CdS: (a) $V_2O_3$ (10 ± 1 nm), (b) $VO_2$ (10 ± 1 nm). The green curve corresponds to resistance vs temperature without light and the red curve with light. In both cases, a small shift in the onset of the transition is labeled.

There are additional major differences between the response of the two CdS/vanadium oxide hybrid heterostructures. Figure 4 shows the resistance versus temperature (R vs T) for the two types of heterostructures, measured with and without light, as a function of time. The green curves show the R vs T before illuminating the samples, the red curves during illumination, and the black dashed curves show the state after turning the light off. The $V_2O_3$ bilayer presents a *volatile* modulation of the MIT (Figure 4 a), i.e. the R vs T recovers its original shape immediately after the light is switched off. This contrasts to the *nonvolatile* modulation in the $VO_2$. The R vs T does not recover its original shape, and the MIT remains suppressed for hours after the light has been turned off. This nonvolatile change, however, is not permanent. The blue curve shows that, after 8 hours at room temperature, the original MIT returns to its initial behavior. Importantly, this demonstrates that the photodoping process does not introduce permanent damage into the sample.

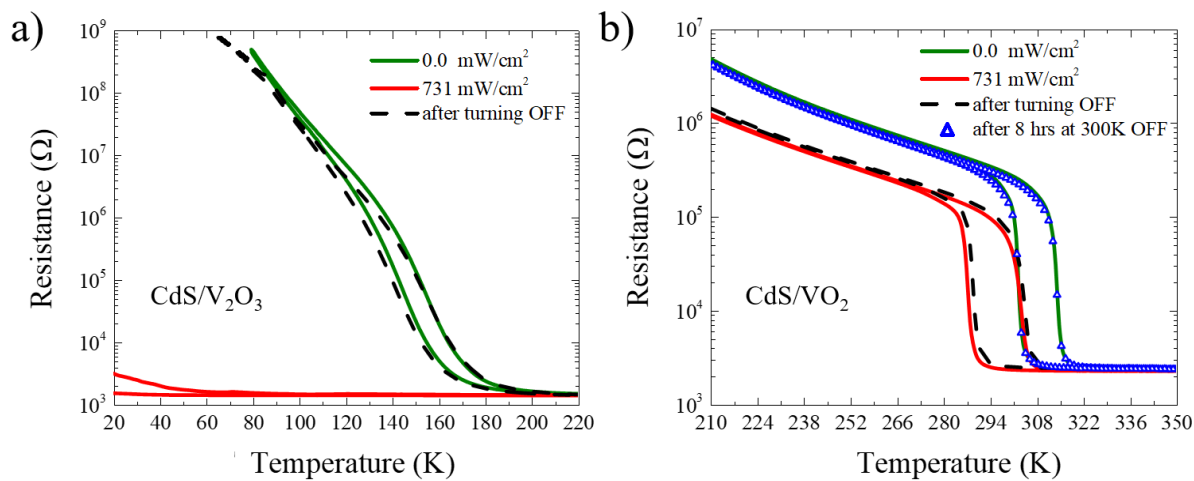

**Figure 4**: **Time dependence of the light-induced modulation.** (a) Volatile changes in the transition temperature of the CdS/V$_2$O$_3$ heterostructure. (b) Non-volatile changes observed in the CdS/VO$_2$ sample. In both cases, the dash lines show the resistance vs temperature immediately after turning off the light. In VO$_2$ case, the blue curve (triangles) shows the recovery of the original state after 8 hrs. at room temperature.

A qualitative explanation of these effects arises by considering the charge carriers in the CdS and vanadates as well as the fact that Mott MITs are highly sensitive to the carrier concentration[31]. It is well documented that conductivity of CdS in the photo excited state is through electron carriers[32]. On the other hand, while in V$_2$O$_3$ the electrical conduction is by holes[33] that of the VO$_2$ is by electrons[34]. The critical carrier concentration needed to undergo a Mott MIT can be affected by adding more photoelectrically created electrons from the CdS, which leak into the vanadate. If the work function of the CdS falls between that of V$_2$O$_3$ and VO$_2$ the observed differences in the effect would have a natural explanation. This would explain qualitatively three important facts discovered in our experiments: 1) the effect is an interface effect, 2) the light completely disrupts the MIT in V$_2$O$_3$ and has a minor effect on the VO$_2$ and 3) the effect is volatile in V$_2$O$_3$ and nonvolatile in VO$_2$.

Alternatively, the differences observed may suggest that the underlying mechanism for the MIT modulation in the various CdS/VOx samples are different. The non-volatile modulation observed in VO$_2$ could be explained by light-induced chemical reactions. On the other hand, CdS is also one of the most prominent photocatalysts[29,35]. Irradiation with photons of energy larger than the band gap may create electron-hole pairs which would produce reduction/oxidation reactions at the CdS/VO$_X$ interface. This would reduce the VO$_2$ into a different Magnéli phase[36]. After the light is switched off, those chemical changes would persist and only slowly reverse over time. The volatile V$_2$O$_3$ case is different, and would be compatible with a modulation of the MIT caused by photodoping. Electrons generated by light in the CdS would transfer across the interface into the ultrathin (hole conducting) $V_2O_3$ layer, doping it and destabilizing the Mott insulating phase.

3. **Conclusion**

In conclusion, we have demonstrated large, light-induced modulation of the MIT in CdS/$VO_X$ heterostructures. For CdS/$V_2O_3$ the $T_{MIT}$ modulation is as large as 140 K, and the resistance can be reduced by six orders of magnitude. For CdS/$VO_2$ the modulation is much smaller, although the resistance can be reduced by as much as a factor of five. The modification is volatile in $V_2O_3$ bilayers, and non-volatile in $VO_2$ bilayers. We show that these effects are caused by the close contact at the CdS/vanadium oxide interface. A simple qualitative explanation suggests that these effects are caused by the differences in the electronic carriers in CdS (electrons), $V_2O_3$ (holes) and $VO_2$ (electrons) and their work functions. An alternative possible explanation for these effects involves photodoping in $V_2O_3$[37], and CdS-mediated photocatalysis in $VO_2$[38,39]. Further experimental work is underway to elucidate the mechanism. The volatile and nonvolatile behavior in two closely related materials systems opens up the possibility for the use of these systems for different neuromorphic applications such as synaptors and neuristors.

In a broader sense, our results show a very promising approach towards the development of novel functionalities in materials using the possible transfer of electronic responses in a properly engineered hybrid heterostructure. This may have further applications as functional materials useful in other optoelectronic applications or systems where a different functionality can affect each other when incorporated into hybrid heterostructures.

4. **Experiment**

We fabricated 10 ± 1 nm thick epitaxial $V_2O_3$ film on top of r-cut sapphire substrates using rf magnetron sputtering from a $V_2O_3$ target, in an 8-mTorr high-purity argon (>99.999%) atmosphere. The substrate temperature during deposition is 720°C, and the sample is cooled at a rate of 80 °C/min after growth. A 10 ± 1 nm thick epitaxial $VO_2$ film was grown by reactive sputtering on top of an $TiO_2$ substrate (oriented along the (001) plane). A 4-mtorr argon/oxygen mix (8% $O_2$) was used during deposition, and the substrate was kept at 600°C during the growth and later cooled down at a rate of 12°C/min (see Figure S.1). CdS 80 ± 1 nm thick film were grown on top of the VOx with rf magnetron sputtering from a CdS target, in a 2-mTorr pure argon atmosphere at 150°C. Electrical transport properties of a CdS film grown directly on $Al_2O_3$ is shown in Figure S.2. In each CdS/VO$_x$ bilayer two Au (40 nm) electrodes were patterned on top of the CdS/VO$_x$ heterostructures films. XRD measurements were done in a Rigaku SmartLab system at room temperature. Single-phase growth is confirmed by XRD, epitaxially along the ⟨012⟩ direction for $V_2O_3$, textured along ⟨002⟩ for $VO_2$, and hexagonal phase direction H⟨002⟩ for CdS. In all samples, negligible changes in the crystal structure were observed upon exposing them to light (see Supplementary information). Transport measurements were carried out on a Montana C2 S50 Cryocooler and TTPX Lakeshore cryogenic probe station, using a Keithley 6221 current source and a Keithley 2182A nanovoltmeter. A white LED was used for the photoconductivity measurements Thorlabs model MCWHLP1.

**Acknowledgements**

This collaborative work was supported as part of the "Quantum Materials for Energy Efficient Neuromorphic Computing" (Q-MEEN-C), an Energy Frontier Research Center funded by the U.S. Department of Energy, Office of Science, Basic Energy Sciences under Award # DE-SC0019273. A R-C thanks the economic support of the mobility research program Salvador de Madariaga from Spanish Ministry of Science.

**Author contributions**

I.K.S. and H. N. conceived the idea. H.N. and J.d.V. designed the experiment. H.N, Y.K. fabricated the samples. H.N performed the transport measurements with assistance from J.d.V, N.M.V., A.R.C., P.L and E.Q. The X-rays diffraction measurements were performed by H.N, C.A, M.H.L. and I.Z. I.Z, O.S, and A.F. carried out the x-ray diffraction analysis. H.N, A.F and I.K.S. wrote the manuscript. All authors participated in the discussion of the results and corrected multiple iterations of the manuscript.

**Competing interests**

The authors declare no competing interests.

**Supplementary Information**: A hybrid optoelectronic Mott insulator

H. Navarro[1*], J. del Valle[1,2], Y. Kalcheim[1], N. M. Vargas[1], C. Adda[1], M-H. Lee[1], P. Lapa[1], A. Rivera-Calzada[3], I. Zaluzhnyy[1], E. Qiu[1], O. Shpyrko[1], M. Rozenberg[4], A. Frano[1] and Ivan K. Schuller[1]

[1] Department of Physics, Center for Advanced Nanoscience, University of California, San Diego, 92093, USA.
[2] Department of Quantum Matter Physics, University of Geneva, 24 Quai Ernest-Ansermet, Geneva, Switzerland.
[3] Departamento de Física de Materiales, Universidad Complutense de Madrid, 28040 Madrid, Spain.
[4] Laboratoire de Physique des Solides, CNRS, Université Paris Saclay, 91405 Orsay Cedex, France.

**Structural characterization of heterostructures**. Figure S.1 show the XRD and XRR for the CdS/$V_2O_3$ and CdS/$VO_2$ bilayers. We observe single-phase epitaxial growth of $V_2O_3$ ⟨012⟩ on the $Al_2O_3$ substrate (r-cut) (see Figure S.1a). We can also see the peaks H⟨002⟩ and with less intensity the peak H⟨102⟩ of the CdS hexagonal structure. The profile fitting of the low-angle X-ray measurements were made with the *Parrat32* code[1], which presents well-defined thicknesses with a roughness of less than 1 nm in each layer. In Figure S.1b we see that the structure of $VO_2$ ⟨002⟩ presents the characteristic peak around 65°, corresponding to ⟨001⟩ orientation and a slightly intense peak of the CdS H⟨110⟩ hexagonal structure of around 44°.

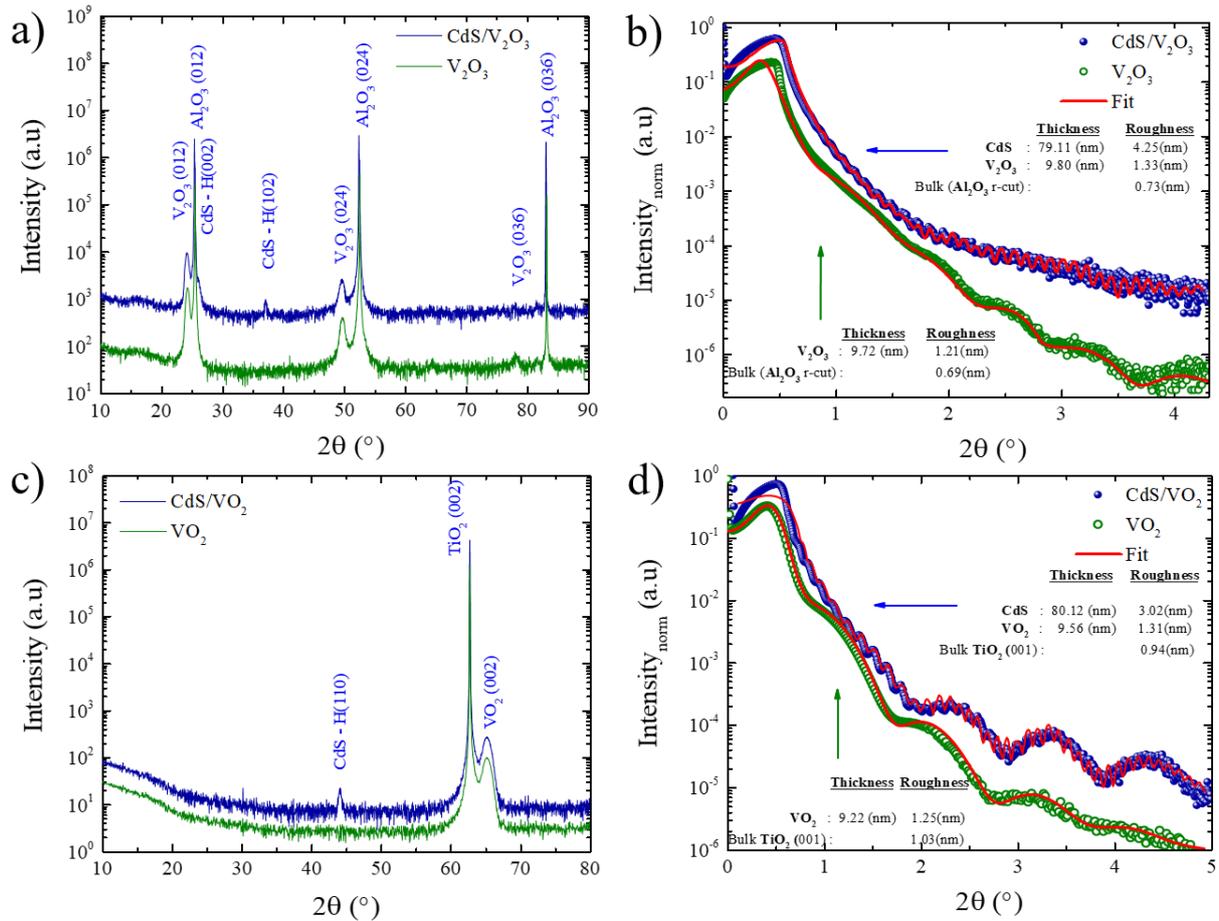

**Figure S.1**: a) XRD pattern for $V_2O_3$ (blue) and for CdS/$V_2O_3$ (red). b) XRR pattern of the CdS/$V_2O_3$ bilayer and the red line correspond to the fitting using the *parrat32* code. c) XRD pattern for the CdS/$VO_2$ (red) bilayers. d) XRR pattern of the CdS/$VO_2$.

Regarding the properties of pure CdS films, an 80 nm film was grown on an $Al_2O_3$ (r-cut) substrate for characterization. The XRD measurements clearly show 2 peaks H⟨002⟩ and H⟨102⟩ associated with the hexagonal structure of the CdS (Figure S. 2a). The electrical transport (Figure S. 2b) was measured in the conventional two-probe configuration. The green curve (darkness) shows the insulating behavior as we decrease the temperature, with a resistance over $10^{12}$ ohms below 200 K. When the CdS is illuminated (red curve) the resistance decreases to $\sim 2\cdot 10^7$ ohms, and remains basically constant down to the lowest temperatures.

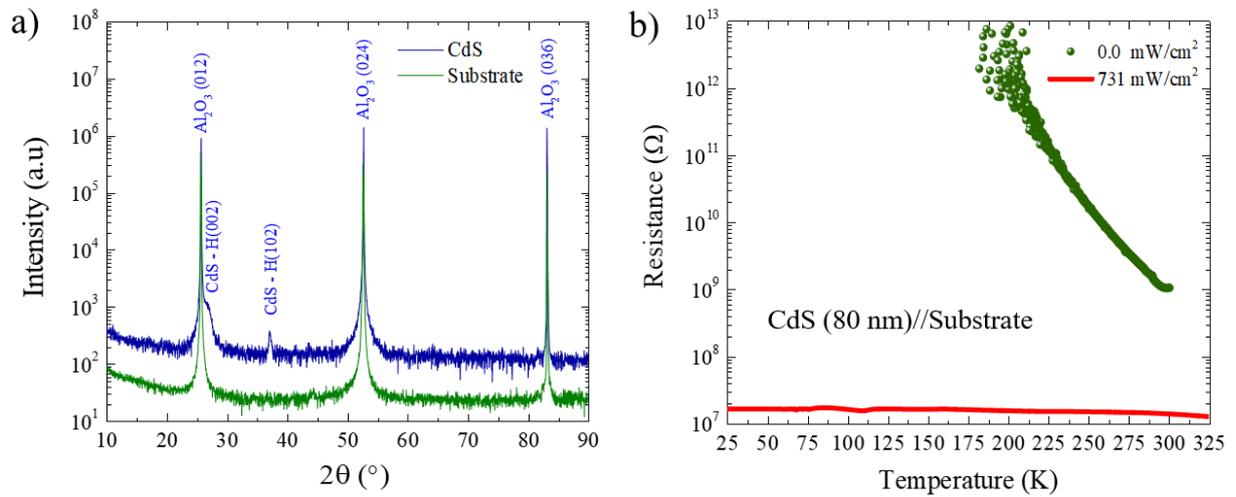

Figure S.2: a) X-rays of the CdS film on Al$_2$O$_3$ (r-cut) substrate. b) Measurements of electrical transport of CdS without light (green) and after being illuminated with P = 731 mW/cm$^2$ (red).

In order to rule out the possibility of heating in the sample, we used a thin strip of platinum on top of the CdS, 50 nm thick and 200 µm wide, as thermometer. The strip resistance was measured as a function of temperature (see Figure S.3) in a conventional four-probe configuration. A shift of 5 K is observed at maximum light power.

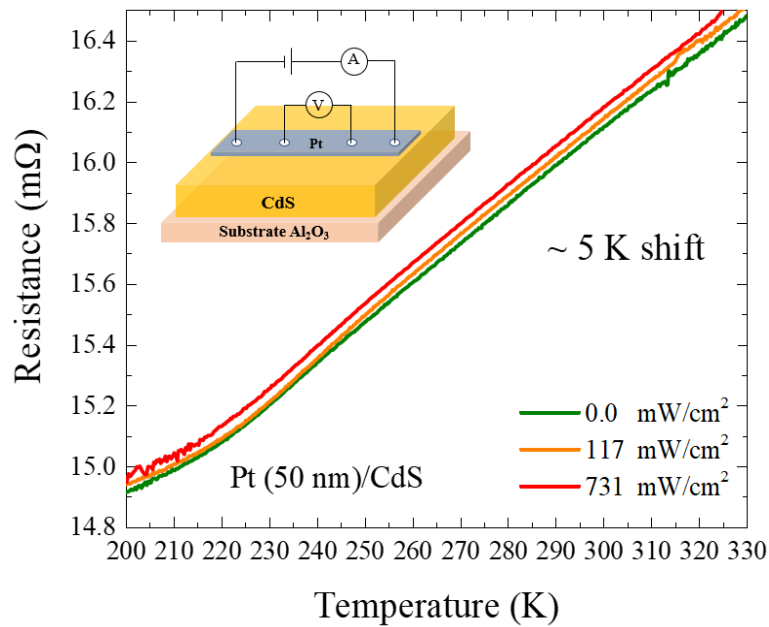

Figure S.3: Measurement of electrical transport of the Pt/CdS bilayer over Al$_2$O$_3$ (r-cut) at different light powers: 0.0 mW/cm$^2$ (green), low power 117 mW/cm$^2$ (red) and at P = 731 mW/cm$^2$ (blue).

It is important to note the difference between a thick sample and a very fine VO$_2$ sample. Clearly, we do not see significant changes in thick VO$_2$ samples compared to a sample 10 times smaller in thickness, as can be seen in Figure S.4.

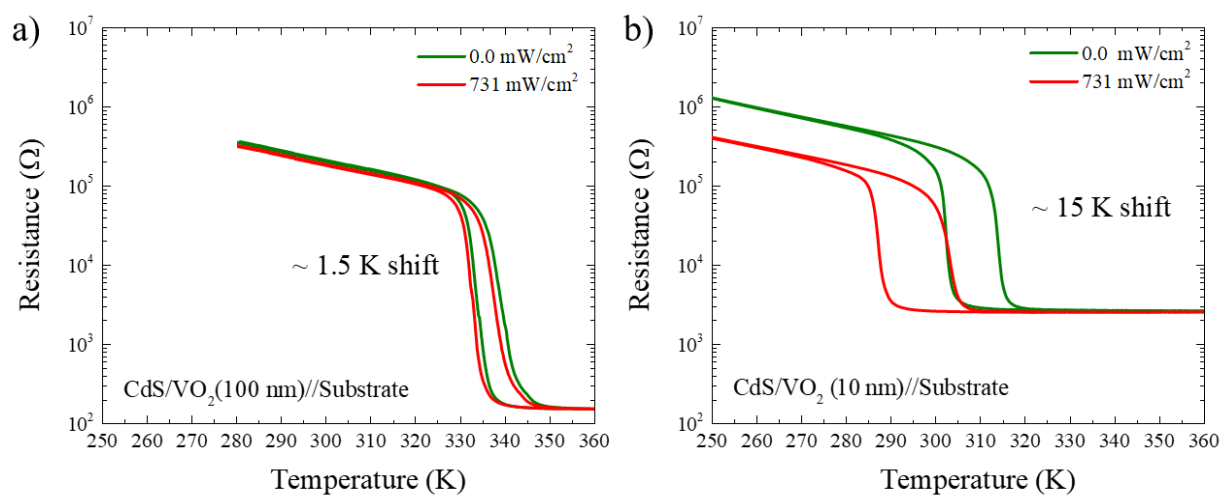

Figure S.4: Behavior in electron transport with and without light of thick and very thin VO$_2$ samples. a) Resistance vs Temperature with film of VO$_2$ (100 nm), b) Resistance vs Temperature with very thin film VO$_2$ (10 nm).

To expand the study of the interface effect localized at the CdS/VO$_X$ boundary, and to understand the causes of T$_{MIT}$ modulation, we have intercalated an Al$_2$O$_3$ insulating barrier (30 nm and 10 nm thick) between CdS (80 nm) and VO$_X$. Measurements without light (green) show MIT for V$_2$O$_3$ (10 nm) and VO$_2$ (7 nm) trilayers. When the samples are illuminated, we don't see a 140 K change in V$_2$O$_3$ as we had without Al$_2$O$_3$. In contrast, the T$_{MIT}$ shift was ~10 K (Fig. S.5 a). Also, in this sample we do see a drop in resistance in the insulating state due to the CdS. On the other hand, for the VO$_2$ trilayer, we do not see a drop in the insulating resistance state, and the T$_{MIT}$ shift is ~4 K. That is, the Al$_2$O$_3$ barrier is sufficient to suppress all the effects seen in Figure 1 in the manuscript, confirming the interfacial and non-thermal nature of this phenomenon.

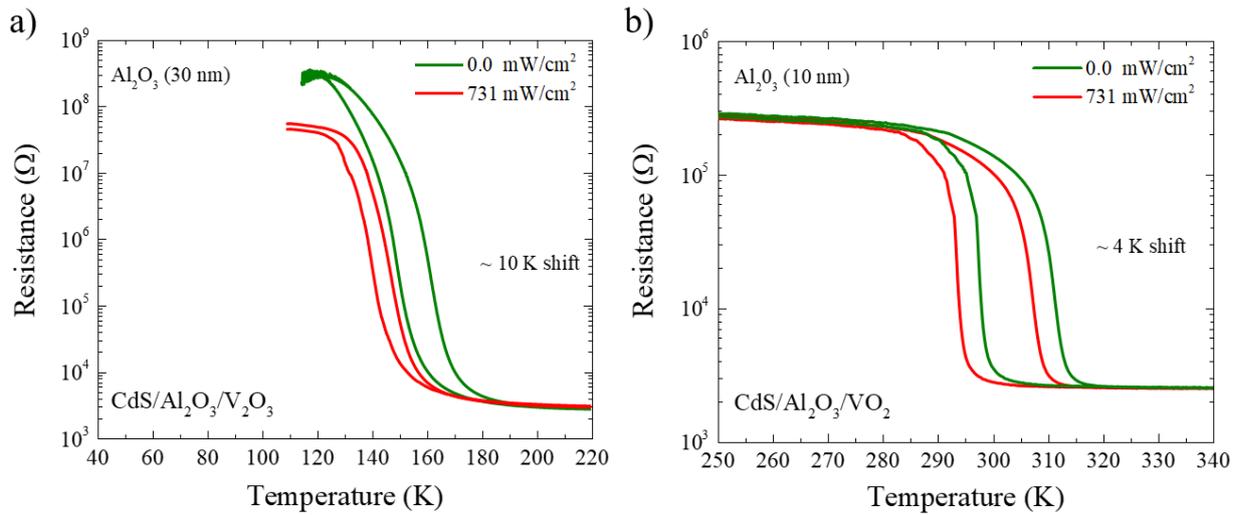

Figure S.5: a) Resistance vs Temperature with $Al_2O_3$ barrier between CdS and $V_2O_3$, b) Resistance vs Temperature with $Al_2O_3$ barrier between CdS and $VO_2$,

One possible explanation for this phenomenology would be heating due to the exposure to light. This, however, would result in a rigid shift of the resistance vs temperature curves, which is not the case since there is also a clear decrease in the insulating state resistance. We also note that the resistance of the CdS alone upon illumination ($>10^7$ Ω – see Figure S.2) is considerably higher than the insulating state resistance of the VOx. Therefore, the photo-induced decrease in resistance below $T_{MIT}$ is a consequence of a reduction of the insulating state resistance of the VOx and not a reduction in the CdS resistance. Another argument against a purely thermal effect is that for the $VO_2$ cases the photo-induced shift in $T_{MIT}$ persists long after the light is turned off (Figure 4), thus ruling out a substantial effect of heating from the light source. This is consistent with direct measurements of the temperature performed using a Pt strip deposited directly on top of the CdS film[2]. From the resistance vs temperature characteristics of the Pt strip (Figure S.3) we estimate a shift of up to 5 K for the maximum applied power density P = 731 mW/cm$^2$. This could explain the small $T_{MIT}$ change, but it is not enough to account for the effects observed in $V_2O_3$ and $VO_2$. Finally, we tested the effect of illuminating bare vanadium oxide films, shown in Figure 3. The shift in the transition temperature is much smaller compared to the CdS/VOx bilayers, and there is no decrease in the insulating state resistance. This implies that the interface with the CdS film plays a fundamental role in the light-induced modification we observe.

**Changes in V₂O₃ and VO₂ crystal lattice under illumination.** The diffraction peaks from the V₂O₃ and VO₂ films were fitted with the following equation

$$I(2\theta) = A + \frac{B}{|2\theta_{sub} - 2\theta|^c} + M \cdot \exp\left(-\frac{(2\theta - 2\theta_{film})^2}{2\sigma^2}\right). \quad (1)$$

Here $A$ represents a constant background, and the tail of a substrate peak is described by a Lorentzian-like function centered at $2\theta_{sub}$ with a magnitude $B$ and exponent $c$. This gives us a good approximation of the substrate peak intensity which decays algebraically as $\propto 1/\theta^c$ away from the peak's maximum. A relatively weak and broad peak from the vanadium oxide film was fitted by Gaussian function centered at $2\theta_{film}$ with a magnitude $M$ and a width $\sigma$. To catch the subtle changes in the vanadium oxide peak position, the fitting procedure was done in two steps:

1) The dark spectrum was fitted with Eq. (1)
2) The background and substrate parameters were fixed ($A, 2\theta_{sub}, B,$ and $c$), and the light spectrum was fitted with only the vanadium oxide-related parameters ($M, 2\theta_{film}$, and $\sigma$) were allowed to change.

Each spectrum was fitted several times, with different starting values of the fitting parameters, and various regions of fitting within $\pm 3°$ around the vanadium oxide peak. To confirm the validity of the procedure we also fitted first the light spectrum, and then the dark spectrum with fixed values of the background and substrate parameters. In each fitting we observed a small but constant shift of the vanadium oxide peak. The obtained parameters for the ⟨012⟩ and ⟨024⟩ peaks from V₂O₃ on Al₂O₃ substrate and the ⟨002⟩ peak from VO₂ on TiO₂ constrain are summarized in Table 1. The example of the fitting is shown in Figs. S.6 and S.7. The mean shift of the diffraction peak was evaluated as an averaged value over N = 10 fitting results with different starting values of the parameters and different regions of fitting

$$\Delta(2\theta) = \frac{1}{N}\sum_{j=1}^{N} \Delta_j(2\theta), \quad (2)$$

where index $j$ enumerates the fitting attempts. The error of the shift was estimated as a root mean square value

$$Error = \sqrt{\frac{1}{N}\sum_{j=1}^{N}\left(\Delta_j(2\theta) - \Delta(2\theta)\right)^2}. \quad (3)$$

The corresponding change of the interplane distance was calculated as it follows from the Bragg's law

$$\frac{\Delta d}{d} = \frac{\Delta(2\theta)}{2\tan\theta}.$$

| $V_2O_3$ | $\langle 2\theta_{V_2O_3}\rangle_{dark}$ | $\langle 2\theta_{V_2O_3}\rangle_{light}$ | $\Delta(2\theta_{V_2O_3})$ | Error of fitting | $\Delta d/d$ (change of interplane distance) |
|---|---|---|---|---|---|
| 012 | 24.1881° | 24.1905° | 0.0024° | 0.0014° | -0.0098% |
| 024 | 49.6307° | 49.6323° | 0.0017° | 0.0006° | -0.0032% |
| $VO_2$ | $\langle 2\theta_{VO_2}\rangle_{dark}$ | $\langle 2\theta_{VO_2}\rangle_{light}$ | $\Delta(2\theta_{VO_2})$ | Error of fitting | $\Delta c/c$ (change of lattice parameter) |
| 002 | 65.2602° | 65.2502° | -0.0101° | 0.001° | 0.014% |

Table S.1: Shift of the diffraction peak and changes in $V_2O_3$ and $VO_2$ crystal lattice under illumination.

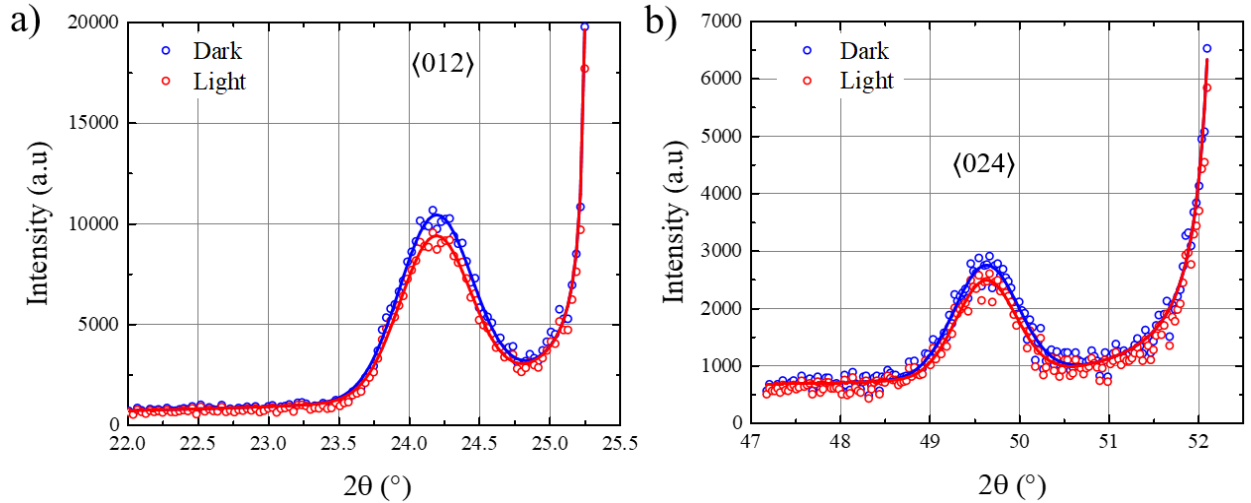

Figure S.6: Fitting of the ⟨012⟩ (left) and ⟨024⟩ (right) diffraction profiles from $V_2O_3$ film without laser illumination (dark) and under laser illumination (light). The strong background peak corresponds to ⟨012⟩ and ⟨024⟩ reflection from the $Al_2O_3$ substrate. Experimental data are shown with circles, and fitting with solid lines.

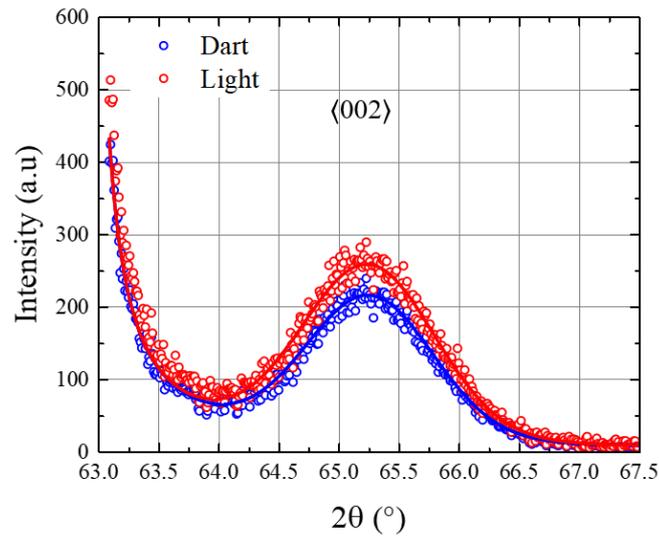

Figure S.7: Fitting of the ⟨002⟩ diffraction profiles from $VO_2$ film without laser illumination (dark) and under laser illumination (light). The strong background peak corresponds to ⟨002⟩ reflection from the $TiO_2$ substrate. Experimental data are shown with circles, and fitting with solid lines.